# Atomic-scale observation of localized phonons at FeSe/SrTiO$_3$ interface

Ruochen Shi[1,2,#], Qize Li[1,2,3,#], Xiaofeng Xu[4,5,#], Bo Han[1,2], Ruixue Zhu[1,2], Fachen Liu[2,6], Ruishi Qi[3], Xiaowen Zhang[1,2], Jinlong Du[2], Ji Chen[7,8,9], Dapeng Yu[2,11], Xuetao Zhu[4,5,10]*, Jiandong Guo[4,5,10]*, Peng Gao[1,2,8,9]*

[1] International Center for Quantum Materials, School of Physics, Peking University, Beijing 100871, China

[2] Electron Microscopy Laboratory, School of Physics, Peking University, Beijing 100871, China

[3] Department of Physics, University of California at Berkeley, Berkeley, CA 94720, USA

[4] Beijing National Laboratory for Condensed Matter Physics and Institute of Physics, Chinese Academy of Sciences, Beijing 100190, China

[5] School of Physical Sciences, University of Chinese Academy of Sciences, Beijing 100049, China

[6] Academy for Advanced Interdisciplinary Studies, Peking University, Beijing 100871, China.

[7] Institute of Condensed Matter and Material Physics, Frontiers Science Center for Nano-optoelectronics, School of Physics, Peking University, Beijing 100871, China.

[8] Collaborative Innovation Center of Quantum Matter, Beijing 100871, China

[9] Interdisciplinary Institute of Light-Element Quantum Materials and Research Center for Light-Element Advanced Materials, Peking University, Beijing 100871, China

[10] Songshan Lake Materials Laboratory, Dongguan, Guangdong 523808, China

[11] Shenzhen Institute for Quantum Science and Engineering (SIQSE), Southern University of Science and Technology, Shenzhen 518055, China

[#] *R. Shi, Q. Li, and X. Xu contributed equally to this work.*

* *Corresponding author. E-mail: xtzhu@iphy.ac.cn, jdguo@iphy.ac.cn, pgao@pku.edu.cn*

# Abstract

In single unit-cell FeSe grown on SrTiO$_3$, the superconductivity transition temperature features a significant enhancement. Local phonon modes at the interface associated with electron-phonon coupling may play an important role in the interface-induced enhancement. However, such phonon modes have eluded direct experimental observations. Indeed, the complicated atomic structure of the interface brings challenges to obtain the accurate structure-phonon relation knowledge from either experiment or theory, thus hindering our understanding of the enhancement mechanism. Here, we achieve direct characterizations of atomic structure and phonon modes at the FeSe/SrTiO$_3$ interface with atomically resolved imaging and electron energy loss spectroscopy in a scanning transmission electron microscope. We find several phonon modes highly localized (~1.3 nm) at the unique double layer Ti-O termination at the interface, one of which (~ 83 meV) engages in strong interactions with the electrons in FeSe based on *ab initio* calculations. The electron-phonon coupling strength for such a localized interface phonon with short-range interactions is comparable to that of Fuchs-Kliewer (FK) phonon mode with long-rang interactions. Thus, our atomic-scale study provides new insights into understanding the origin of superconductivity enhancement at the FeSe/SrTiO$_3$ interface.

Single unit-cell (UC) FeSe grown on SrTiO$_3$ substrate has attracted strong research interest for its remarkably high superconductivity transition temperature, which is about an order of magnitude higher compared to that of bulk FeSe[1–3]. The anomalously large superconducting gap occurring only in the first UC of FeSe indicates that superconductivity is substantially enhanced by the existence of interface[1,4,5]. Replica bands were firstly observed in angle-resolved photoemission spectroscopy (ARPES) experiments, which have an approximately 90-to-100-meV energy shift between replica bands and main bands, despite some debate[6], were identified as the signature of electron-phonon coupling[7–10]. Recent high resolution electron energy loss spectroscopy (HREELS) experiments suggest that the phonons involved in electron-phonon coupling are likely to be the Fuchs-Kliewer (F-K) phonons of SrTiO$_3$[11–13]. The similarity between energy of the F-K phonon and energy shift of replica bands indicates that this phonon could contribute predominately to the electron-phonon coupling.

The emergence of localized phonons at the interface is widely understood to be a consequence of the breakdown of translational symmetry, which alters the local bonds and, subsequently, the lattice vibrations at the interface. However, to precisely probe the highly localized phonons across the FeSe/SrTiO$_3$ interface is challenging for the surface analysis techniques such as ARPES and HREELS due to their large probe size and their unique setup configurations. On the other hand, the lack of accurate knowledge on the atomic structure of interface in previous studies blurs our understanding of the structure-property relation. It can be expected that the interface properties, including the electronic structures, phonon modes and electron-phonon coupling strongly depend on the atomic structure. In fact, the atomic structure of this interface is sensitive to sample history such as annealing and surface treatment during sample preparation[14,15]. Theoretically, previous investigations either approximated the phonon structure without providing an accurate depiction of the atomic structure[16–18] (relying solely on the normal monolayer Ti-O termination, which however does not align well with experimental observations[14,19,20]), or they concentrated primarily on the electron bands at the interface[21–23]. The variable and complicated atomic structure of interface poses challenges for *ab initio* calculations aiming to meticulously reproduce or predict interfacial properties such as the phonon structure or electron-phonon coupling. To date, the localized phonons across the FeSe/SrTiO$_3$ interface are still largely unknown, not to mention precisely correlating with their atomic arrangements or superconductivity properties, which motivates our present study.

In this work, we study FeSe/SrTiO$_3$ interface by using scanning transmission electron microscopy-electron energy loss spectroscopy (STEM-EELS). The cutting-edge developments in STEM-EELS have made it possible to directly image phonon excitations at the nanoscale credit to its high spatial and energy resolution[24–28], which is very suitable for the study of interfacial phonons in heterostructures[29–33]. Meanwhile, the ability to reveal the atomic structure, electronic state, and phonon mode allows us to understand their interrelationships and thus the underlying mechanism.

We first combine high-angle annular dark-field (HAADF) images with atomically resolved core-loss spectra to recognize the atomic structure of double layer Ti-O termination at FeSe/SrTiO$_3$ interface. The structure reconstruction exists in the top layer Ti-O (Top layer is defined as layer adjacent to FeSe). From the atomically resolved phonon spectra across the interface, highly localized interfacial phonon modes can be observed. The *ab initio* calculations of interfacial phonon by taking subsistent double Ti-O termination layer into consideration further confirm that the interfacial phonons originated from the double layer Ti-O termination, i.e., phonons with energy ~18 meV and ~81 meV are centered at the top layer and phonons with energy ~51 meV and ~80 meV are centrally enhanced at the bottom layer. Particularly, we identify a strongly coupling interfacial (SCI) phonon mode at the double layer Ti-O termination. This mode can promote pronounced electron-phonon coupling at the interface, whose strength is even comparable to that of the previously reported Fuchs-Kliewer (FK) phonon mode. Thus, such a localized interface phonon likely plays a pivotal role in the enhancement of interface superconductivity. Our atomic-scale measurements of phonons across the FeSe/SrTiO$_3$ interface and studies of correlation among the atomic structure, phonon structure and quantum properties help us to understand the past experiments and provide new insights for the mechanism of the enhanced interfacial superconductivity.

In our study, the substrate treatment, sample growth and annealing procedure are exactly the same as previously reported conditions that are optimal for superconductivity[34]. We first study the atomic structure of FeSe/SrTiO$_3$ interface by atomically resolved STEM-HAADF and core-level EELS. Fig. 1a is a HAADF image of FeSe/SrTiO$_3$ interface viewed along [100] zone axis. The double layer Ti-O termination is observed, in accordance with former researches[14,19,20]. The bottom/top layers of double layer Ti-O termination are marked in the Fig. 1a with purple/orange arrows, and named as Ti-O(B)/Ti-O(T) in the following text. The existence of double layer Ti-O termination can also be confirmed from the atomically resolved core-loss EELS of Fe-$L_{2,3}$ edge (upper panel) and Ti-$L_{2,3}$ edge (lower panel) in Fig. 1b. Changes of Ti-$L_{2,3}$ edge at Ti-O(B) and Ti-O(T) layers are attributed to the distortion of $TiO_6$ octahedron and possible oxygen vacancies introduced during sample annealing[35,36](see Extended Data Fig. 1 for better visualization). In the Ti-O(T) layer, there is an extra atom contrast between ordinary top Ti sites (red arrow in Fig. 1a), which is expected to be oxygen atom position and invisible in the HAADF image. This has been reported and explained as the reconstructed Ti-O termination layer[23], e.g., recent studies revealed $\sqrt{13} \times \sqrt{13}$ R33.7° reconstruction of Ti-O layer after FeSe was grown[20,22]. In fact, various $SrTiO_3$ surface reconstructions with additional Ti-O layer have been reported[37–39]. To verify the assumption that the extra atom contrast comes from reconstruction, we performed the relaxation with density functional theory (DFT) on $FeSe/SrTiO_3$ structure without reconstruction, and with $\sqrt{2} \times \sqrt{2}$ R45°, $\sqrt{5} \times \sqrt{5}$ R26.6°, $\sqrt{10} \times \sqrt{10}$ R18.3°, $\sqrt{13} \times \sqrt{13}$ R33.7° reconstruction, and corresponding STEM-HAADF image simulations using QSTEM[40] (see Extended Data Fig. 2 for details). From the simulation, structure without reconstruction does not show extra atom contrast on red-arrow-pointed site. Although the $\sqrt{2} \times \sqrt{2}$ R45° structure does show extra atom contrast, the distance between bottom Se layer and top Ti layer is significantly smaller than the experimental result. The other three structures are sub-structures of $\sqrt{13} \times \sqrt{13}$ R33.7° reconstruction[39], and their simulated images all agree well with the experiment. Due to the reported experimental evidence of $\sqrt{5} \times \sqrt{5}$ R26.6° reconstruction under similar annealing conditions[41] (see Methods for detail), we pick $\sqrt{5} \times \sqrt{5}$ R26.6° reconstruction as a rational structure to explain our experimental result. Its simulated HAADF image and atomistic models are shown in Fig. 1c.

We then measure phonon spectra across the $FeSe/SrTiO_3$ interface by atomically resolved STEM-EELS. The HAADF image of the acquisition region and a line profile of corresponding EEL spectra are shown in Fig. 2a and Fig. 2b respectively. The clear contrast between Sr column and Ti-O column demonstrates sufficiently high spatial resolution to distinguish interfacial phonon and the rationality of column-by-column spectra analysis. As shown in Fig. 2b, the $SrTiO_3$ transverse optical (TO) branch phonon around 65 meV splits into two branches as approaching the interface from $SrTiO_3$ side, a transformation attributable to the emergence of the interfacial phonon. One of the phonon branches experiences a blue-shift, positioning itself within the energy gap of 65-100 meV, while the other transitions towards lower energy domains. For better visualization, the spectra extracted from bulk $SrTiO_3$ (blue), Ti-O(B) (purple), Ti-O(T) (orange), FeSe layer adjacent to the interface ($FeSe_{int}$, yellow) and bulk FeSe (green) are shown in Fig. 2c. Their spatial regions are labeled by dashed rectangles with corresponding color in Fig. 2a. We find that several local interfacial phonon modes emerge due to the presence of double layer Ti-O termination. Even with broadening over adjacent atomic layers, we have still successfully pinpointed the localization centers of these modes. The red arrows in Fig. 2c point to the spectral features show enhancement centered at Ti-O(T) layer, whose energies are ~18 meV and ~81 meV. The black arrows point to the features that are centrally located at Ti-O(B) layer whose energies are ~51 meV and ~80 meV.

To better separate the intrinsic spectra from interface, we applied non-negative matrix factorization (NMF) on the whole spectrum image. NMF is a powerful tool to provide well-interpretable characteristic of data, which has already been successfully applied to the identification of EEL spectra[42,43]. Three components are found to best describe the acquired data. Their intensity maps are shown in Fig. 2d-f. Fig. 2g shows the line profile of their intensity map across the interface. Component I shows atomic contrast consistent with Ti column in HAADF and decays fast as approaching the interface, indicating its origin from the vibration of Ti-O plane in $SrTiO_3$. Component III shows contrast of Sr atom and FeSe column, thus is attributed to the sum of vibration signals from FeSe and Sr atom in $SrTiO_3$, considering their similarity in vibration energy. Intriguingly, component II is highly localized at the interface, constructing a peak centered at the Ti-O(T) layer, spreading over the double layer Ti-O termination and the first FeSe layer. This helps to explain the large enhancement of superconductivity only occurring in the first UC of FeSe. The full-width-at-half-maxima (FWHM) of component II (w in Fig. 2g) is ~1.3 nm. A different and independent approach to extract the features of interfacial phonons is finding the minimum difference between measured spectrum and possible linear combination of two bulk spectra. The result is shown in Extended Data Fig. 3, in which the width ~1.3 nm agrees well with the NMF result, further confirming the presence and highly localized nature of interfacial phonons.

Notably, the feature above 45 meV in FeSe spectra still remains even far from the interface, exceeding the max frequency of pure FeSe[17]. Therefore, it must come from SrTiO$_3$ substrate. This can be found in NMF spectrum of component III as well (Extended Data Fig. 4a). We ascribe these signals to the phonon polaritons (PPs), i.e., the F-K phonons, of SrTiO$_3$, which are found to penetrate into FeSe in previous works[11–13]. To confirm this, we fitted the spectra of NMF component III, and compared it to the result that was acquired under on-axis experimental setup. The similarity of spectrum shape and fitted energy between on-axis and off-axis results supports our assumption (see Extended Data Fig. 4b-d for detail).

To get further insights into the interfacial phonon, we carried out *ab initio* phonon calculations on an interface model of FeSe on the double-layer Ti-O terminated SrTiO$_3$ (see Methods). Corresponding to the experimental spectra, the projected phonon density of states (PDOS) of bulk SrTiO$_3$, Ti-O(B) layer, Ti-O(T) layer, FeSe$_{int}$ layer and bulk FeSe are plotted in Fig. 3a. Similarly, the spectrum features ~18 meV and ~81 meV that are significantly enhanced at Ti-O(T) layer (red arrows), and spectrum features ~51 meV and ~80 meV that are enhanced at Ti-O(B) layer (black arrows), are spotted. The eigenvectors of these modes from side view and top view are illustrated in Extend Data Fig. 5a-d. These modes involve the vibrations much stronger at either Ti-O(B) or Ti-O(T) layer, indicating their highly localized nature (more details in Extended Data Fig. 5e-h). The projected dispersions of interface models further demonstrate the localized nature of interfacial phonons and dynamical stability of the reconstructed structure (Extended Data Fig. 6), i.e., significant imaginary frequency only exists in the reconstruction-free interface model. All these characters agree well with the experimental findings, despite a small mismatch of energy that is within the accuracy of the DFT calculations.

To establish a direct correlation between observed interfacial phonons and enhanced superconductivity, we extracted the electron-phonon coupling features from the calculations. The phonon linewidth due to electron-phonon coupling mapped on the phonon dispersion and the Eliashberg spectral function at Γ point for calculated structure is shown in Fig. 3b. Particularly, we find that a SCI phonon mode which has the strongest coupling strength in our calculations. Firstly, it has an energy of ~83 meV and maximum line width among all modes at Γ point, which corresponds to the peak of the same energy in the Eliashberg spectral function. From the Eliashberg spectral function curves, we obtain the electron-phonon coupling constant[44] of the SCI phonon mode as $\lambda \approx 0.10$, a value comparable with the previously reported electron-phonon coupling strength from the forward scattering model[8,45]. Secondly, its eigenvector from side and top view is shown in Fig. 3c, suggesting this mode is caused by the out-of-phase vibration of Ti and O ions. Extended Data Fig.7 demonstrates the localized nature of this SCI mode. Such vibrations induce a dipole, and change the electric field at the interface, leading to interactions with electrons in FeSe[7]. Thirdly, the electron-phonon coupling usually leads to the phonon softening[46]. To confirm this behavior in our system, we performed calculations with a double layer Ti-O terminated $SrTiO_3$ surface model (without FeSe). Distinctively, we find a mode analogous to SCI mode with respect to vibration eigenvector (see Extended Data Fig. 8). This mode has an energy of ~93 meV at the FeSe-free $SrTiO_3$ surface, which is about 10 meV higher than the energy of the SCI mode. The pronounced softening in energy for the same vibrational mode can solely be attributed to the presence of FeSe. The substantial phonon softening in this system implies the presence of strong electron-phonon coupling induced by the observed SCI mode, i.e., the phonons of $SrTiO_3$ strongly interact with the electrons in FeSe.

In previous studies, F-K phonons from the $SrTiO_3$ substrate have usually been regarded as the primary origin of interfacial electron-phonon coupling in this system[11–13]. The intense long-range dipolar field generated by the substrate F-K phonons has been widely believed to enhance the electron pairing in FeSe, although the exact microscopic interaction mechanism is still awaiting clarification. Our results show that besides to the F-K phonons with long-range interactions, the localized interfacial phonons with short-range interactions also make unique contributions to the electron-phonon coupling, providing new insights into understanding the underlying mechanisms of enhanced superconductivity at interfaces.

In summary, we carried out direct measurements of phonon spectra at FeSe/$SrTiO_3$ interface and correlated the features with the unique double layer Ti-O termination with reconstruction. We implemented the *ab initio* calculations of interfacial phonons by taking subsistent double Ti-O termination layer into consideration, which agree well with the experiment. We find that highly localized phonons are emergent at the interface, which promote intense electron-phonon coupling. These findings take an essential step towards revealing the role of interface in the interface-enhanced superconducting systems.

## Methods

**Sample growth and TEM sample preparation.** A FeTe(20 UC) /FeSe(20 UC)/SrTiO$_3$ structure was grown by molecular beam epitaxy. Before the epitaxial growth of FeSe films, the Nb-doped (0.5 wt %) SrTiO$_3$ (001) substrates (from Shinkosha Co. Ltd) were pretreated in ultrahigh vacuum (UHV) at 1000 °C for 45 min to obtain a Ti-O plane terminated surface. The high-quality FeSe film was grown by co-depositing high-purity Fe (99.99%) and Se (99.99+%) with a flux ratio ~1:10 onto treated SrTiO$_3$ held at 470 °C. After growing FeSe, the sample was annealed at 480°C for 3 hours under UHV. Afterwards, FeTe capping layer was grown by co-depositing Fe and Te with a flux ratio ~1:5 onto FeSe held at 385 °C. The cross-sectional TEM sample was prepared by ThermoFisher Helios G4 UX Focused Ion Beam (FIB) system.

**TEM characterization and EELS data acquisition.** The STEM-HAADF images shown in Fig. 1a, Extended Data Fig. 1b and Extended Data Fig. 2a were recorded using an aberration-corrected FEI Titan Themis G2 operated at 300 kV. The beam convergence semi-angle was 30 mrad and the collection angle was 39-200 mrad. The HAADF image shown in Fig. 2a was recorded at 60 kV using a Nion U-HERMES200 microscope equipped with both a monochromator and aberration correctors, with 35-mrad convergence semi-angle and 80-200 mrad collection semi-angle. All the EELS data were acquired on the same Nion U-HERMES200 microscope operated at 60 kV, with 35-mrad convergence semi-angle and 24.9-mrad collection semi-angle. EELS data shown in Fig. 1b, Extended Data Fig. 1a, Extended Data Fig. 1c and Extended Data Fig. 4c-d was collected under the on-axis experimental setup, namely the center of the collection aperture was placed at the center of the direct beam disk. Data shown in Fig. 1b and Extended Data Fig. 1a was recorded with 128×12 pixels within the range of 6 nm across the interface. The energy dispersion was set as 0.262 eV/channel. Data shown in Extended Data Fig. 4c-d was collected with 128×16 pixels within the range of 8 nm across the interface. The energy dispersion was 0.5 meV/channel. Data shown in Fig. 2b-g, Extended Data Fig. 3 and Extended Data Fig. 4a-b was collected under the off-axis experimental setup, in which the electron beam was displaced from optical axis along the [010] direction of SrTiO$_3$ with 60 mrad to greatly reduce the contribution from the long-range dipole scattering[28,47]. The data was collected with 80×10 pixels within the range of 8 nm across the interface. The energy dispersion was 0.5 meV/channel.

**EELS data processing.** All acquired EEL spectra were processed by custom-written MATLAB code. The EEL spectra were first aligned by their normalized cross-correlation. Subsequently, the spatial drift-correction was applied for obtaining line-scan data. The background of the core-loss EELS data shown in Fig. 1b and Extended Data Fig. 1a-c was fitted by power-law and then subtracted from the whole space. The phonon spectra shown in Fig. 2b-g, Extended Data Fig. 3 and Extended Data Fig. 4 were multiplied by the square of energy rather than fitting with a background function to resolve the difficulty of fitting the background function at low energy, and to treat spectra in different spatial components uniformly. This method has already been performed effectively in processing EELS data acquired in SrTiO$_3$ and other analogous materials[33]. Lucy–Richardson deconvolution was then employed to ameliorate the broadening effect caused by the finite energy resolution. NMF was performed to decompose the off-axis data. NMF is a computational method that decomposes a non-negative matrix into the product of two non-negative matrices, often used for feature extraction[48]. We found 3 components best describe our data. The spectra in Extended Data Fig. 4 was fitted by gaussian peak. In Extended Data Figure 3, the interface component of the spectra was extracted by fitting measured spectra with linear combination of SrTiO$_3$ spectrum and FeSe spectrum. The fitting was performed by minimizing $\|S(\omega) - a_1 S_{\text{SrTiO}_3}(\omega) - a_2 S_{\text{FeSe}}(\omega)\|$ while keeping the residual non-negative, where $S(\omega)$ is the measured spectrum in Fig. 2b, $S_{\text{SrTiO}_3}$ means the bulk SrTiO$_3$ spectra, $S_{\text{FeSe}}$ means the bulk FeSe spectra, and $a_1$, $a_2$ are adjusted coefficients.

***Ab initio* calculations.** Density functional theory calculations were performed using Quantum ESPRESSO[49,50] with the Perdew-Burke-Ernzerhof for solid (PBEsol) exchange-correlation functional[51] and the projector augmented wave (PAW) method pseudopotential[52]. The kinetic energy cut-off was 60 Ry for wavefunctions and 600 Ry for charge density and potential. The reconstruction-free, $\sqrt{2} \times \sqrt{2}$ R45° reconstruction, $\sqrt{5} \times \sqrt{5}$ R26.6° reconstruction, $\sqrt{10} \times \sqrt{10}$ R18.3° reconstruction and $\sqrt{13} \times \sqrt{13}$ R33.7° reconstruction FeSe/SrTiO$_3$ slab structures containing 1 UC FeSe connected to the double-layer Ti-O terminated 3 UC SrTiO$_3$ were built, for which an in-plane lattice constant of $a_{\text{SrTiO}_3} = 3.893$ Å (the optimized lattice constant of bulk SrTiO$_3$ under the used exchange-correlation functional and pseudopotential) and an out-of-plane lattice constant of $c = 40$ Å was chosen. The total atom numbers of these structures are 22, 45, 110, 217 and 277, respectively. The lattice mismatch between SrTiO$_3$ and FeSe was ignored. All the structures were optimized while keeping lattice constant invariant until the residual force was below 10$^{-5}$ Ry/Bohr on every atom and the total energy gradient below 10$^{-10}$ Ry, reaching the numerical limit of the software.

For the phonon calculation, we used FeSe (1 UC)/SrTiO$_3$(3 UC) interface models. Only reconstruction-free and $\sqrt{5} \times \sqrt{5}$ R26.6° reconstruction FeSe/SrTiO$_3$ structures are applied, because the $\sqrt{2} \times \sqrt{2}$ R45° reconstruction structure is not consistent with our experimental observation, and $\sqrt{10} \times \sqrt{10}$ R18.3° reconstruction or $\sqrt{13} \times \sqrt{13}$ R33.7° reconstruction structures contain too many atoms to be calculated in DFT framework. The dynamical matrices and force constants were obtained using Phonopy[53]. The treatment of non-analytical term is implemented in Quantum Espresso package. The projected phonon DOS (PDOS) was calculated by interpolating the dynamical matrix on a 15×15×5 q-mesh. No evident change in PDOS can be observed in the SrTiO$_3$ 2 UC away from the interface compared with bulk SrTiO$_3$, manifesting that our model is large enough to distinguish the interfacial SrTiO$_3$ form the bulk SrTiO$_3$. The calculated PDOS contains small imaginary frequencies occupying ~1% of the total PDOS of SrTiO$_3$ along with the double-layer Ti-O termination, close to that obtained from calculation in bulk SrTiO$_3$ performed in the same way. But the imaginary frequencies did not appear in the region around the Γ point, thus not confounding the analysis of phonon modes at the Γ point. The eigenvectors in Fig. 3c, Extended Data Fig. 5, 7 and 8 are picked at Γ point. As a comparison, a surface structure only containing 3 UC double layer Ti-O terminated SrTiO$_3$ was relaxed separately and then used for phonon calculation. A higher-energy surface mode ~93 meV was found, whose eigenvector resembles that of the SCI mode. As another comparison, phonon calculation of fully relaxed reconstruction-free FeSe/SrTiO$_3$ structure was performed using both FD method with a 2×2×1 supercell and DFPT method with a 2×2×1 q-mesh. No difference is found between results of two methods. Its projected phonon dispersion is shown in Extended Data Fig. 6b. Several phonon modes with large imaginary frequencies emerge at Γ point, indicating the dynamical instability of the structure.

To confirm the spatial characteristics of the SCI mode, we examined all phonon modes in the interface model containing 3 UC $SrTiO_3$. The calculation result also shows the presence of this SCI mode, which only involves atoms of the top layer of $SrTiO_3$, consolidating that this mode is highly localized at the interface (see Extended Data Fig. 7).

For the phonon calculation of bulk $SrTiO_3$ and bulk FeSe in Fig.3a, we built the conventional one-unit-cell $SrTiO_3$ and FeSe, and fully relaxed with constrain of in-plane lattice constant $a = 3.893$ Å. The interatomic forces are calculated by DFPT with a 4×4×4 q-mesh. The PDOS was calculated by diagonalizing dynamical matrix interpolated on a 20×20×20 q-mesh, and projected onto all the atoms in the unit cell.

For the electron-phonon coupling calculation, a $\sqrt{5} \times \sqrt{5}$ R26.6° reconstructed FeSe/$SrTiO_3$ slab structures containing 1 UC FeSe connected to the double-layer Ti-O terminated 1 UC $SrTiO_3$ (60 atoms in total) was applied using density functional perturbation theory (DFPT) implemented in Quantum ESPRESSO. We chose 1 UC $SrTiO_3$ instead of 3 UC (110 atoms in total) for electron-phonon coupling calculation due to the limitation of computing resource. As a comparison, phonon structure corresponding to this model was also calculated using finite displacement (FD) method with a combination of Phonopy and Quantum ESPRESSO using a 1×1×1 supercell. No evident change in phonon structure was observed in results of DFPT or FD method for this structure. A dense mesh of 16×16×4 k-points was used for the sum of electron–phonon coefficients at the Fermi energy. Other parameters are kept same as in the above text.

## Acknowledgement

The work was supported by National Natural Science Foundation of China (52125307, 11974023, 52021006, T2188101), the "2011 Program" from the Peking-Tsinghua-IOP Collaborative Innovation Center of Quantum Matter, Youth Innovation Promotion Association, CAS. The sample growth at CAS was supported by the National Key R&D Program of China (2017YFA0303600, 2021YFA1400200), the National Natural

Science Foundation of China (11874404, 11974399, 11974402), and the Strategic Priority Research Program of Chinese Academy of Sciences (XDB33000000). X.T.Z. was partially supported by the Youth Innovation Promotion Association of Chinese Academy of Sciences. We acknowledge Prof. Peter Rez at Arizona State University for helpful discussion. We acknowledge Electron Microscopy Laboratory of Peking University for the use of electron microscopes. We acknowledge High-performance Computing Platform of Peking University for providing computational resources for the DFT and FD calculation.

## Data availability

The data that support the findings of this study are available from the corresponding author upon request.

## Code availability

Custom MATLAB codes used for data processing and DFT related post-processing are available from the corresponding author upon request.

## Author Contribution

R.C.S., Q.Z.L. and X.F.X. contributed equally to this work. P.G. and J.D.G. conceived the project; X.F.X. grew the sample with the guidance of X.T.Z. and J.D.G.; R.C.S. performed the STEM-EELS experiment and data analysis assisted by Q.Z.L., B.H., F.C.L., R.S.Q., and X.W.Z. with the guidance of P.G.; Q.Z.L. and R.C.S. performed *ab initio* calculations with the guidance of J.C.; R.X.Z. prepared the TEM sample. B.H. acquired the atomic-resolution STEM-HAADF image. X.F.X., X.T.Z. and J.D.G. helped the data interpretation. R.C.S. wrote the manuscript with the help of Q.Z.L. under the direction of X.T.Z., J.D.G. and P.G.; All the authors contributed to this work through useful discussion and/or comments to the manuscript.

## Competing Interests

The authors declare no competing interests.

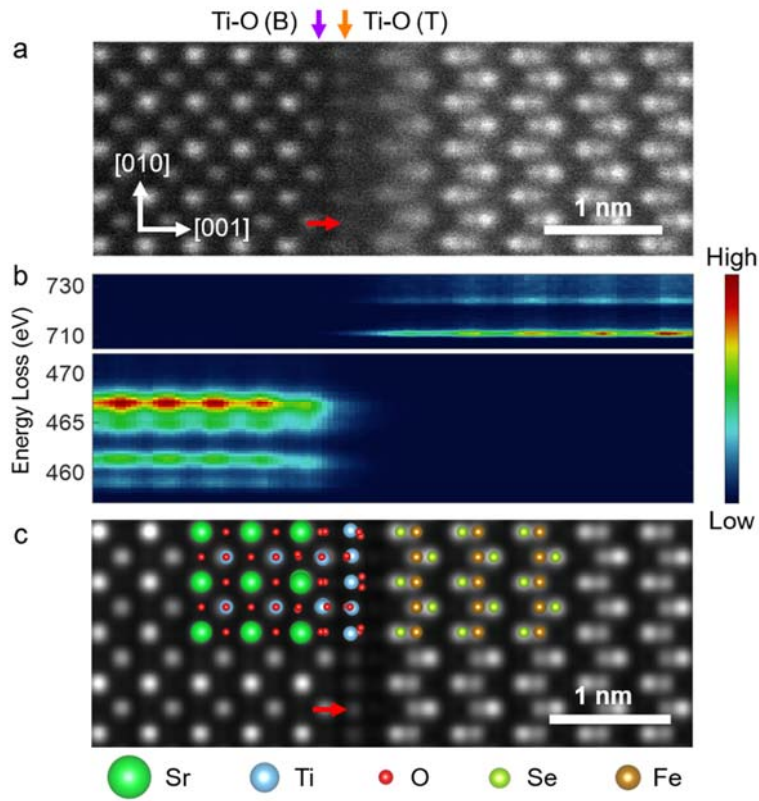

**Fig. 1 | The atomic structure of FeSe/SrTiO₃ interface**. **a.** A HAADF image of FeSe/SrTiO$_3$ interface from [100] zone axis. The purple and the orange arrows indicate the Ti-O(B) and Ti-O(T) layers respectively. **b.** The core-loss of Fe-$L_{2,3}$ edge (upper panel) and Ti-$L_{2,3}$ edge (lower panel) across the interface. **c.** The simulated HAADF image and overlaid atomistic model. The red arrows in **a** and **c** point to the extra atom contrast between ordinary top Ti site.

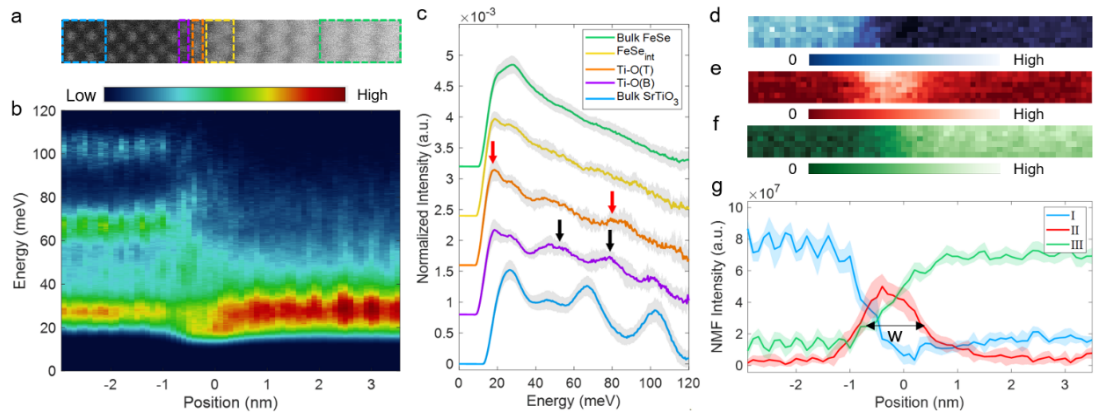

**Fig. 2 | The atomically resolved phonon spectra at FeSe/SrTiO$_3$ interface. a.** The HAADF image showing the region where EELS spectra are acquired. **b.** The line profile of atomically resolved off-axis EELS spectra across the interface. **c.** The spectra extracted from bulk SrTiO$_3$ (blue), Ti-O(B) (purple), Ti-O(T) (orange), FeSe layer adjacent to the interface (yellow) and bulk FeSe (green). Their spatial regions are labeled by dashed rectangles with corresponding color in **a**. The red and black arrows point to the main features that are enhanced at Ti-O(T) and Ti-O(B) layer, respectively. The gray shades are the standard deviations. **d – e.** The NMF intensity map for component I **(**Ti-O in SrTiO$_3$**, d**), component II **(**interface**, e**) and component III **(**FeSe and Sr in SrTiO$_3$, **f**). **g.** The line profile of NMF map across the interface. The colored shades are corresponding standard deviations. The FWHM for component II is labelled as w in the figure.

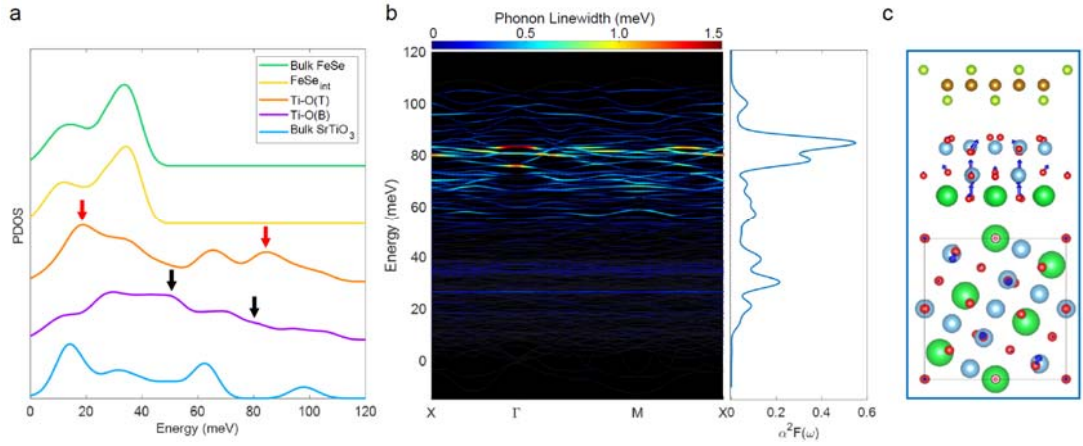

**Fig. 3 | The calculated phonon DOS, dispersion and electron-phonon coupling for FeSe/SrTiO$_3$ interface. a.** The calculated PDOS extracted from bulk SrTiO$_3$ (blue), Ti-O(B) in interface structure(purple), Ti-O(T) in interface structure (orange), FeSe layer in the interface structure (yellow) and bulk FeSe (green). The red and black arrows point to the main features that are enhanced at Ti-O(T) and Ti-O(B) layer, respectively. **b.** Phonon linewidths due to electron-phonon coupling mapped on phonon dispersion and the Eliashberg spectral function at Γ point for calculated structure. The phonon with energy ~83 meV has the strongest coupling strength with FeSe electrons at Γ point, i.e., SCI mode. **c.** The side view (upper panel) and top view (lower panel) of phonon eigenvectors of SCI mode in calculated interface structure.

**Extended Data Figures**

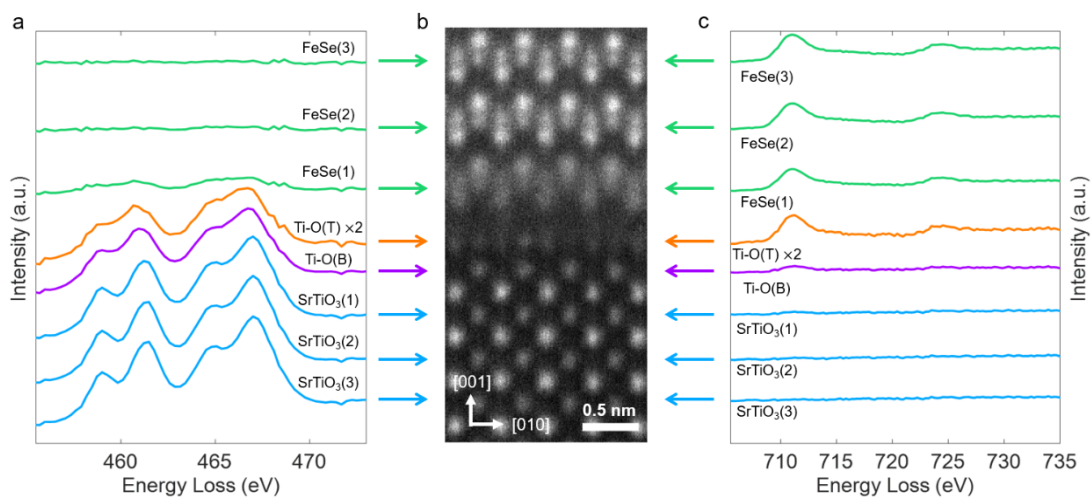

**Extended Data Fig. 1 | The core-loss spectra of Ti-$L_{2,3}$ edge and Fe-$L_{2,3}$ across the FeSe/SrTiO$_3$ interface. a.** The core-loss spectra of Ti-$L_{2,3}$ edge extracted from SrTiO$_3$ that is 1-3 UC away from the interface (blue), Ti-O(B) (purple), Ti-O(T) (orange), and FeSe that is 1-3 UC away from the interface (green). The spectra are offset, and the spectra of Ti-O(T) is multiplied by 2 for clarity. Four peaks representing the $t_{2g}$-$e_g$ splitting of Ti-$L_{2,3}$ edges are clearly visible. The peaks are broader and the splitting is less pronounced at Ti-O(B) and Ti-O(T) layers, indicating a local change of electron states at the interface. **b.** The HAADF image showing the region where the core-loss data are acquired. The color arrows point to the atom layers corresponding to the region where the spectra are extracted. **c.** The core-loss spectra of Fe-$L_{2,3}$ edge extracted from SrTiO$_3$ that is 1-3 UC away from the interface (blue), Ti-O(B) (purple), Ti-O(T) (orange), and FeSe that is 1-3 UC away from the interface (green). The spectra are offset, and the spectra of Ti-O(T) is multiplied by 2 for clarity. We note that the Fe core-loss signal is less localized in the Fe columns next to the Se columns. The possible reasons include the roughness of STO surface (atomic step of surface along the observation direction), the intrinsic delocalization effect, or the mixing of Fe-Se.

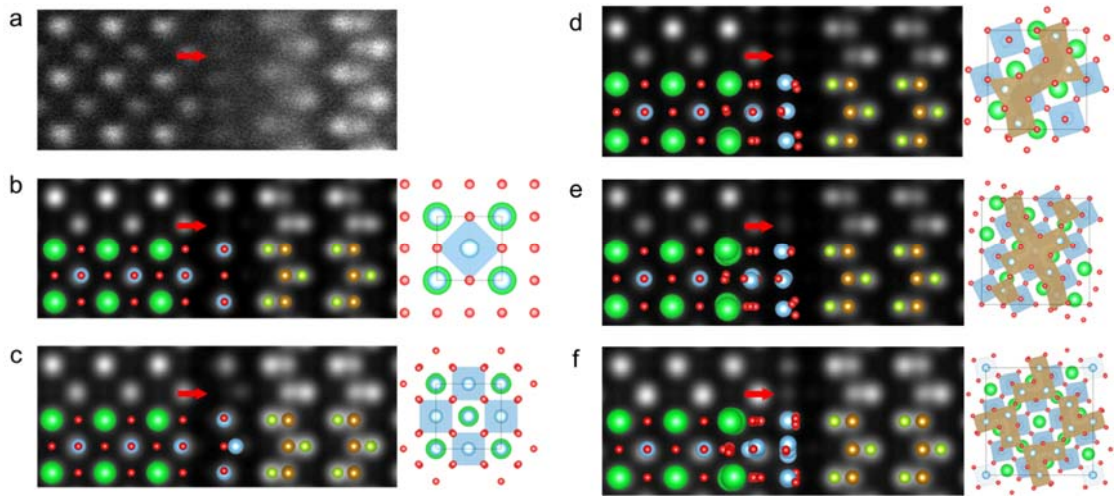

**Extended Data Fig. 2 | The experimental and simulated HAADF images of various FeSe/SrTiO₃ interface structures. a.** Enlarged experimental HAADF image of FeSe/SrTiO₃ interface viewed from [100] zone axis. **b – f.** Simulated HAADF image overlaid by atomistic model from side view (left panel) and atomistic model from top view (right panel) of SrTiO₃ surface with no reconstruction (**b**), $\sqrt{2} \times \sqrt{2}$ R45° reconstruction (**c**), $\sqrt{5} \times \sqrt{5}$ R26.6° reconstruction (**d**), $\sqrt{10} \times \sqrt{10}$ R18.4° reconstruction (**e**) and $\sqrt{13} \times \sqrt{13}$ R33.7° reconstruction (**f**). The blue polyhedrons in top views are complete $TiO_6$ octahedrons, while brown polyhedrons are $TiO_5$ tetrahedrons. The red arrow in each panel point to the extra atom contrast between ordinary top Ti site.

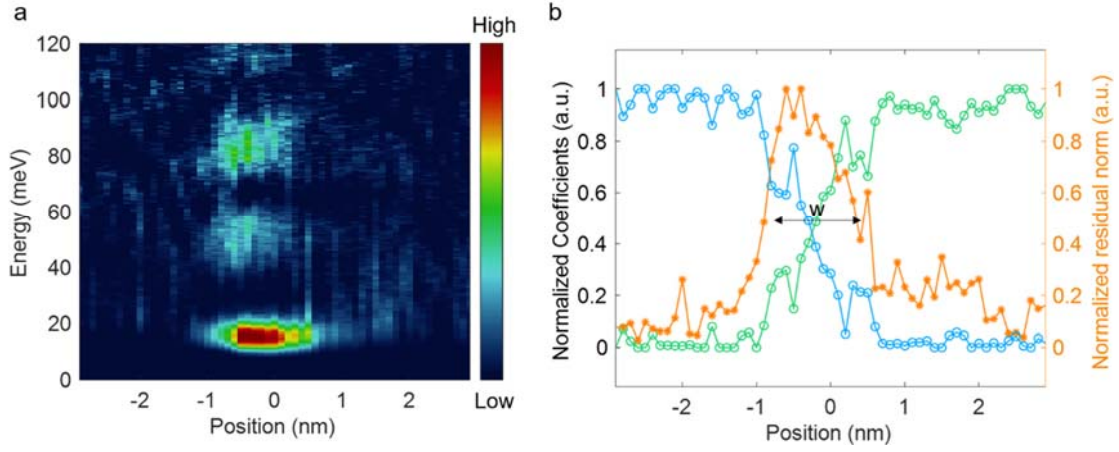

**Extended Data Fig. 3 | The interface component of the spectra extracted by fitting measured spectra with linear combination of SrTiO₃ spectrum and FeSe spectrum.** The fitting was performed by minimizing $\|S(\omega) - a_1 S_{\text{SrTiO}_3}(\omega) - a_2 S_{\text{FeSe}}(\omega)\|$ while keeping the residual non-negative, where $S(\omega)$ is the measured spectrum (Fig. 2b), $S_{\text{SrTiO}_3}$ means the bulk SrTiO₃ spectra, $S_{\text{FeSe}}$ means the bulk FeSe spectra, and $a_1$, $a_2$ are adjusted coefficients. **a.** The fitting residual, which represents the intensity of newly emergent interfacial phonons. The match between fitting result and NMF result proves the self-consistency of the data and analysis process. **b.** The line profile of normalized fitting coefficients from bulk SrTiO₃ (blue) and bulk FeSe (green), and the normalized residual norm (orange). Again, the FWHM of residual norm w is ~1.3 nm, in excellent agreement with the result of NMF.

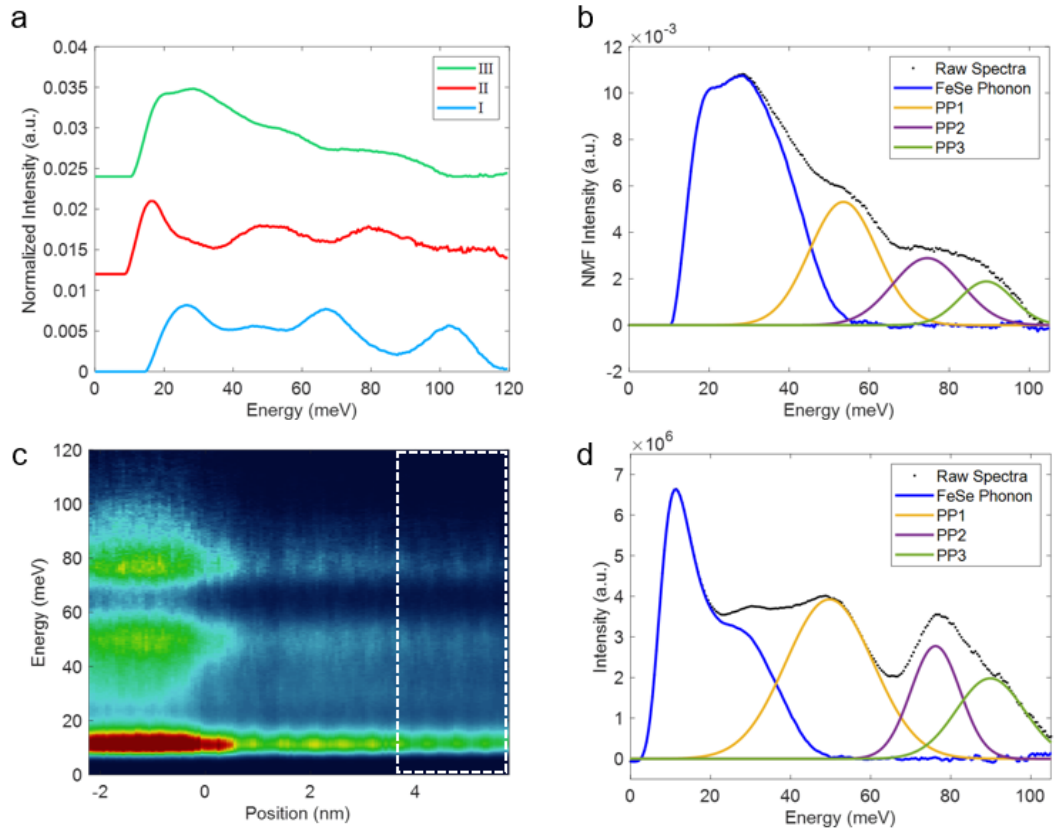

**Extended Data Fig. 4 | The off-axis NMF spectra, on-axis EELS spectra and decomposition of phonon polariton. a.** The NMF spectra for component I, II and III performed on the off-axis data. **b.** The gauss peak fitting for NMF component III spectrum. The spectral features above 45 meV are fitted by three gaussian peaks which are attributed to three branches of phonon polaritons (PPs). These signals were collected even in off-axis experimental geometry via multiple scattering, i.e., the high-energy electron undergoes inelastic scattering by PP and elastic scattering by Bragg scattering subsequently. The large overlap of diffraction disk in 35 mrad illumination enables this process easily. The peak energies are ~53 meV for PP1 (yellow), ~74 meV for PP2 (purple) and ~89 meV for PP3 (green). **c**. The line profile of atomically resolved on-axis EEL spectra across the interface. The spectral features above 45 meV extend to whole FeSe region without energy shift or significant intensity decay due to the delocalized nature of PPs. **d.** The gauss peak fitting for spectrum extracted from white dashed rectangle in **c**. Similar to **b**, three gaussian fitted PPs are located at energy ~50 meV (PP1), ~76 meV (PP2), and ~90 meV (PP3).

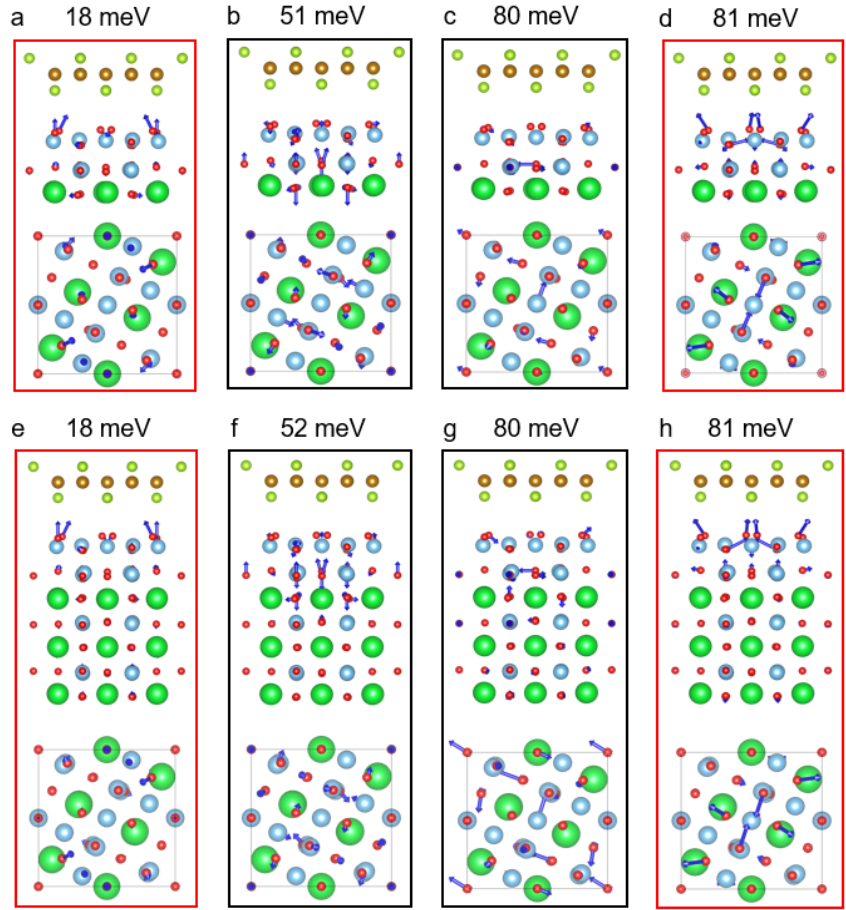

**Extended Data Fig. 5 | The phonon eigenvectors of interfacial modes for FeSe/SrTiO₃ interface. a-d.** The side view (upper panel) and top view (lower panel) of phonon eigenvectors in 1 UC interface structure with energy ~18 meV (**a**), ~51 meV (**b**), ~80 meV (**c**) and ~81 meV (**d**). **e-h.** The corresponding phonon eigenvectors in 3 UC interface structure with energy ~18 meV (**e**), ~52 meV (**f**), ~80 meV (**g**) and ~81 meV (**h**). The colors of the boxes correspond to the colors of the black and red arrows in Fig. 3a. These modes involve the vibrations much stronger at either Ti-O(B) or Ti-O(T) layer in both interface models, indicating their highly localized nature.

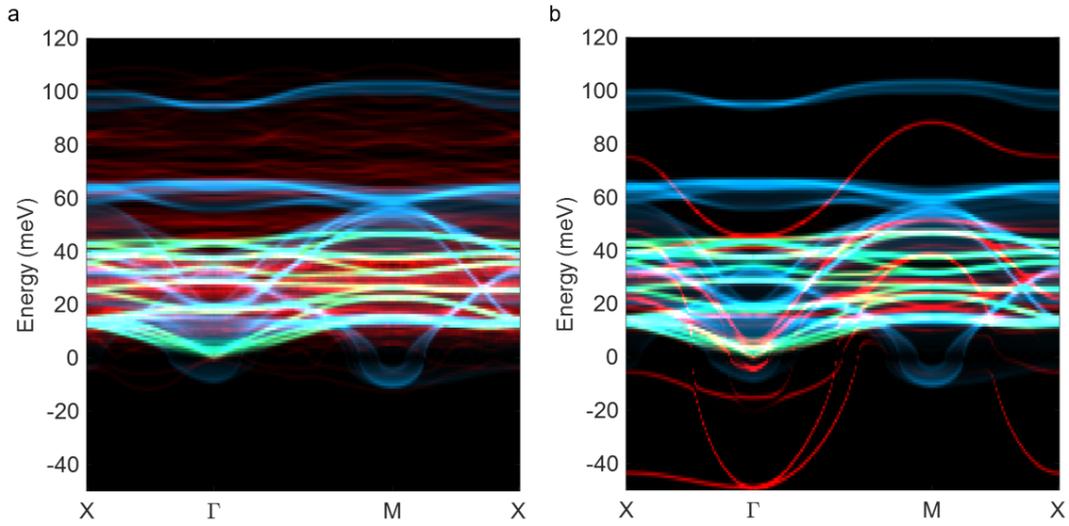

**Extended Data Fig. 6 | The projected phonon dispersion of bulk SrTiO$_3$, bulk FeSe and FeSe/SrTiO$_3$ interface.** The green and blue lines are phonon dispersion in bulk FeSe and bulk SrTiO$_3$ respectively, projected along [001] direction. The red lines are phonon dispersion projected onto interfacial Ti-O atoms in **a.** $\sqrt{5} \times \sqrt{5}$ R26.6° reconstructed 1 UC interface model and **b.** reconstruction-free 3 UC model. The red lines in **a** are flat, and lying outside the projected bands of bulk materials, further confirming the localized nature of interfacial phonons. Besides, no interfacial phonon with imaginary frequency is found at Γ point and only one interfacial phonon with no more imaginary frequency than that of bulk SrTiO$_3$ is found at other points in momentum space in **a**. On the contrary, several interfacial phonons with very large imaginary frequencies at Γ point are found in **b.** This indicates the interface with $\sqrt{5} \times \sqrt{5}$ R26.6° reconstruction is dynamically more stable than the reconstruction-free one.

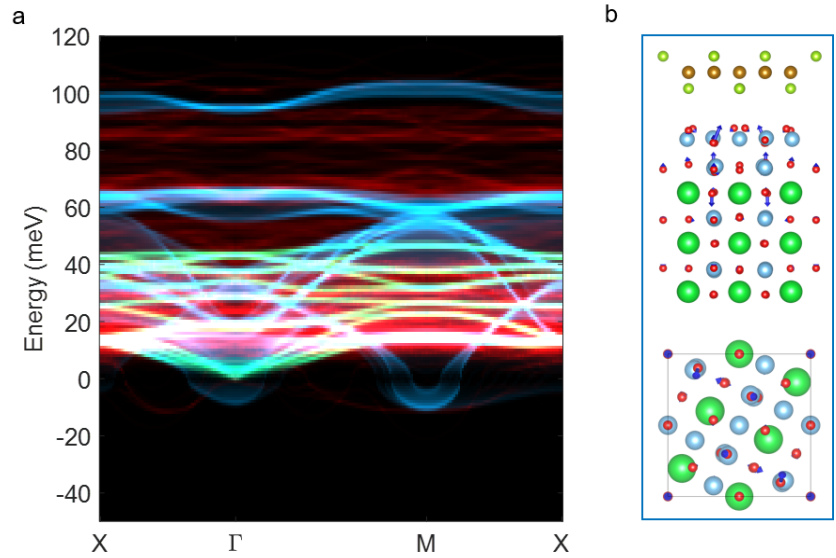

**Extended Data Fig. 7 | The projected phonon dispersion and eigenvectors of interfacial modes in FeSe (1 UC)/SrTiO$_3$(3 UC) interface model. a.** The green and blue lines are phonon dispersion in bulk FeSe and bulk SrTiO$_3$ respectively, projected along [001] direction. The red lines are phonon dispersion projected onto interfacial Ti-O atoms. **b.** The side view (upper panel) and top view (lower panel) of phonon eigenvectors corresponding to the mode in Fig. 3c. The dispersion in which red lines are lying outside the projected bands of bulk materials, and the phonon eigenvector only involving atoms of the top layer of SrTiO$_3$ help confirm the localized nature of interfacial phonons.

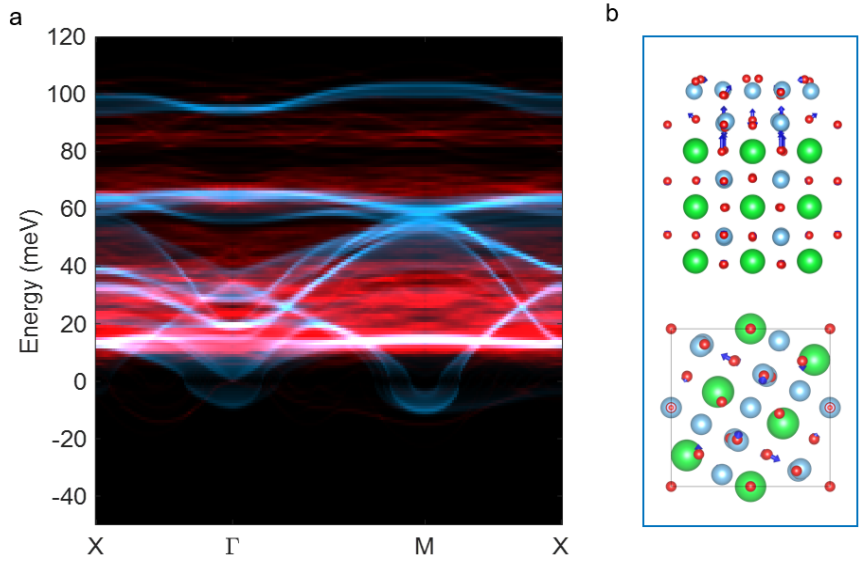

**Extended Data Fig. 8 | The projected phonon dispersion and eigenvectors of interfacial modes in 3 UC SrTiO$_3$ surface model. a.** The blue lines are phonon dispersion in bulk SrTiO$_3$ projected along [001] direction. The red lines are phonon dispersion projected onto surface Ti-O atoms. **b.** The side view (upper panel) and top view (lower panel) of phonon eigenvectors corresponding to the mode in Fig. 3c. This mode has an energy of ~93 meV, which is about 10 meV higher than the energy of the SCI mode. The pronounced softening in energy for the same vibrational mode can solely be attributed to the presence of FeSe.